\documentstyle[prd,aps,amssymb,preprint,epsfig]{revtex}
\input prepictex
\input pictex
\input postpictex

\begin{document}


\def\ubar{{\bar u}}
\def\qbar{{\bar q}}
\def\nbar{{\bar n}}
\def\Lambdabar{{\bar\Lambda}}

\def\calM{{\cal M}}
\def\calO{{\cal O}}
\def\calL{{\cal L}}
\def\calP{{\cal P}}

\def\vslash{v\hspace{-1.8mm}/}
\def\nslash{n\hspace{-2.1mm}/}
\def\Aslash{A\hspace{-2.5mm}/}
\def\Dslash{D\hspace{-2.5mm}/}
\def\nbarslash{\nbar\hspace{-2.1mm}/}
\def\calPslash{\calP\hspace{-2.5mm}/}


\def\etal{{\it et al.}}
\def\ibid#1#2#3{{\it ibid.} {\bf #1} (#2) #3}
\def\epjc#1#2#3{Eur. Phys. J. C {\bf #1} (#2) #3}
\def\ijmpa#1#2#3{Int. J. Mod. Phys. A {\bf #1} (#2) #3}
\def\jhep#1#2#3{J. High Energy Phys. {\bf #1} (#2) #3}
\def\mpl#1#2#3{Mod. Phys. Lett. A {\bf #1} (#2) #3}
\def\npb#1#2#3{Nucl. Phys. {\bf B#1} (#2) #3}
\def\plb#1#2#3{Phys. Lett. B {\bf #1} (#2) #3}
\def\prd#1#2#3{Phys. Rev. D {\bf #1} (#2) #3}
\def\prl#1#2#3{Phys. Rev. Lett. {\bf #1} (#2) #3}
\def\rep#1#2#3{Phys. Rep. {\bf #1} (#2) #3}
\def\zpc#1#2#3{Z. Phys. {\bf #1} (#2) #3}
\def\ibid#1#2#3{{\it ibid}. {\bf #1} (#2) #3}


\title{Heavy-to-light form factors of $B$ decays at large recoil}
\author{Jong-Phil Lee\footnote{e-mail: jplee@phya.yonsei.ac.kr}}
\address{Department of Physics and IPAP, Yonsei University, Seoul, 120-749, Korea}

\tighten
\maketitle

\begin{abstract}

The form factors of $B\to\pi(\rho)$ decays are analyzed using the light-cone
sum rules in the framework of the soft-collinear effective theory (SCET).
We establish the sum rules for the leading and the next-to-leading order (NLO)
nonperturbative functions of the SCET.
Explicit calculation shows that (leading)+(NLO) $\sim 1/E^2+1/E^3$, where $E$
is the large recoiling energy.
The results are compatible with the literatures.
Also the validity of this hybrid formalism is discussed.

\end{abstract}

\pacs{}
\pagebreak

\section{Introduction}

One of the least known elements of the Cabibbo-Kobayashi-Maskawa (CKM) matrix
$V_{ub}$ is the highlight of recent studies in particle physics.
$B$ factories in KEK and SLAC are now producing copious $B$ mesons, reducing 
experimental errors considerably.
In spite of remarkable developments in theory part during the last decade,
theoretical uncertainties in determining $|V_{ub}|$ are still in the way of
higher accuracy;
very recent measurement by $BABAR$ is 
$|V_{ub}|=(3.64\pm0.22\pm0.25^{+0.39}_{-0.56})\times 10^{-3}$ from the rare
decays $B\to\rho e\nu$ \cite{BABAR}.
\par
The extraction of $|V_{ub}|$ greatly depends on the heavy-to-light form factors.
As for heavy-to-heavy transitions, the heavy quark effective theory (HQET) based
on the heavy quark symmetry (HQS) works well to simplify the relations of the
relevant form factors \cite{HQET}.
Semileptonic charmed decay modes $B\to D^{(*)}\ell\nu$ are good stuffs to apply
the HQET and to extract $|V_{cb}|$ \cite{HQET,Vcb}.
\par
If the final state hadron is light, however, constraints from the HQET on the
heavy-to-light form factors are less strong.
In the kinematical regions where the light hadron carries a large energy
$E\sim m_b$, other effective theories have been developed such as large energy
effective theory (LEET) \cite{LEET} or soft collinear effective theory (SCET)
\cite{SCET}.
We adopt the SCET as a basic framework in this analysis.
\par
In SCET, an energetic light particle defines the light-cone direction $n^\mu$
whose component of the momentum is the largest and scales as $E$.
A small expansion parameter $\lambda$ is introduced to be the ratio of the
transverse momentum $p_\perp$ to $E$.
The smallest component of the momentum is the backward one which is of order 
$\lambda^2 E$.
A systematic expansion of $\lambda$ is possible in SCET, and the subleading
analysis was already done in \cite{Chay}.
\par
The heavy-to-light currents of the full QCD are matched to the effective 
currents of SCET below the scale $\mu\lesssim E$.
Effective weak currents are composed of the collinear quarks and the heavy
quark fields, and a new set of Wilson coefficients is introduced during the
matching.
\par
It was already known that three independent functions suffice to describe
heavy-to-light transition at large recoil \cite{Charles}.
The SCET analysis reproduced the same features in \cite{SCET}, and it was found
that more functions are needed at subleading order of $\lambda$ \cite{Chay}.
\par
The authors of \cite{Charles} argued that the form factors of heavy-to-light
decays scale as $1/E^2$, and cross-checked by the light-cone sum rule (LCSR)
calculations.
The argument for $1/E^2$ is the same as \cite{Ball0}, where the end-point
configuration for the "soft contribution" to $B\to\pi(\rho)$ yields 
$\calO(1/m_b^2)$ behavior of the form factors, up to the normalization 
convention.
Very similar scaling was first derived in \cite{Chernyak}.
In this paper, we give the HQET-LCSR calculations in the framework of SCET.
LCSR is among the most reliable nonperturbative methods, especially for 
$B\to\rho$ decays \cite{Ali}.
In the traditional QCD sum rules nonperturbative nature is encoded in the
vacuum condensates.
But the concept of vacuum condensate may cause unphysical behavior of the
distribution amplitudes at the end point region \cite{Ball}.
The LCSR is free from this nuisance because the low-energy physics is 
parametrized by the well-defined distribution amplitudes.
Recently, the HQET-based LCSR analysis is given for $B\to\pi(\rho)$ in 
\cite{Wang}.
The usefulness of this hybrid formalism was discussed in \cite{Liu}, and the
next-to-leading order of $1/m_Q$ calculations were given.
The authors of \cite{Liu} found that the large recoil limit of their results
agrees with the LEET analysis of \cite{Charles}.
\par
The fact that both SCET and LCSR are adequate to describe the heavy-to-light
decays at large recoil is the main motivation of this paper.
We combine in this work the HQET LCSR \cite{Wang} with subleading SCET 
\cite{Chay} to examine the energy scalings of the form factors.
The $1/E^2$ scaling of leading functions is checked and the energy dependence
of the form factors are analyzed up to $\calO({\lambda})$.
In this work, we neglect the hard gluon-spectator effects which might occur
for simplicity.
\par
In Sec.\ II, $B\to\pi(\rho)$ transition matrix elements are parametrized in the
context of SCET at order $\lambda$.
Section III is devoted to evaluate the LCSR within the SCET framework.
Relevant SCET functions are expressed in terms of distribution amplitudes for
the light mesons.
Our results and discussions appear in Sec.\ IV and summary is given in Sec.\ V.

\section{Matrix elements in the SCET}

The standard parametrization of $B\to\pi(\rho)$ decay matrix elements is
\begin{mathletters}
\begin{eqnarray}
&&\langle\pi(p)|V^\mu|B(p_B)\rangle=
 f_+(q^2)\left[p_B^\mu+p^\mu-\frac{m_B^2-m_\pi^2}{q^2}q^\mu\right]
 +f_0(q^2)\frac{m_B^2-m_\pi^2}{q^2}q^\mu~,\\
&&\langle\rho(p,\epsilon^*)|V^\mu|B(p_B)\rangle=
 \frac{2V(q^2)}{m_B+m_\rho}i\epsilon^{\mu\nu\alpha\beta}\epsilon^*_\nu p_\alpha
   (p_B)_\beta~,\\
&& \lefteqn{
\langle\rho(p,\epsilon^*)|A^\mu|B(p_B)\rangle}\nonumber\\
&=&
 i(m_B+m_\rho)A_1(q^2)\epsilon^{*\mu}
 -i\frac{A_2(q^2)}{m_B+m_\rho}\epsilon^*\cdot p_B (p_B+p)^\mu
 -i\frac{A_3(q^2)}{m_B+m_\rho}\epsilon^*\cdot p_B q^\mu~,
\end{eqnarray}
\end{mathletters}
where $V^\mu(A^\mu)=\ubar\gamma^\mu b(\ubar\gamma^\mu\gamma_5 b)$, $q=p_B-p$,
and $\epsilon^{*\mu}$ is the polarization vector of $\rho$.
If the final-state mesons were heavy, then the HQS would allow to relate all
the form factors, leaving only one independent Isgur-Wise function.
For energetic light mesons, a similar reduction of the form factors can be
obtained in the SCET systematically in powers of $\lambda$.
\par
First we briefly review the structure of the SCET, following \cite{SCET,Chay}.
In the $B$-rest frame, the energetic light particle defines the light-cone 
direction $n^\mu=(1,0,0,1)$.
The quark momentum $p$ has large part ${\tilde p}$ and small one $k$,
\begin{equation}
p={\tilde p}+k~,~~~
{\tilde p}\equiv(\nbar\cdot p)\frac{n}{2}+p_\perp~,
\end{equation}
where $\nbar^\mu=(1,0,0,-1)$ and 
$p^\mu=(p^+, p^-, p_\perp)=(n\cdot p, \nbar\cdot p, p_\perp)$.
Extracting the large momentum ${\tilde p}$ defines a new quark field $q_{n,p}$
\begin{equation}
q(x)=\sum_{{\tilde p}}e^{-i{\tilde p}\cdot x}q_{n,p}(x)~,
\end{equation}
from which the collinear quark fields are introduced:
\begin{equation}
\xi_{n,p}=\frac{\nslash\nbarslash}{4}q_{n,p}~.
\end{equation}
The effective Lagrangian can be constructed at leading ($\calL_0$) and 
subleading ($\calL_1$) order of $\lambda$ as \cite{Chay}
\begin{eqnarray}
\calL_0&=&{\bar\xi}_n\Bigg\{n\cdot(iD-gA_n)
 +(\calPslash_\perp-g\Aslash^\perp_n)W\frac{1}{{\bar\calP}}W^\dagger
 (\calPslash_\perp-g\Aslash^\perp_n)\Bigg\}\frac{\nbarslash}{2}\xi_n~,
\nonumber\\
\calL_1&=&{\bar\xi}_n\Bigg\{i\Dslash_\perp W\frac{1}{{\bar\calP}}W^\dagger
 (\calPslash_\perp-g\Aslash^\perp_n)
 +(\calPslash_\perp-g\Aslash^\perp_n)W\frac{1}{{\bar\calP}}W^\dagger 
 i\Dslash_\perp\Bigg\}\frac{\nbarslash}{2}\xi_n~,
\end{eqnarray}
where
\begin{equation}
W=\Bigg[\exp\Bigg(\frac{1}{{\bar\calP}}g\nbar\cdot A_n\Bigg)\Bigg]~,
W^\dagger=\Bigg[\exp\Bigg(g\nbar\cdot A^*_n\frac{1}{{\bar\calP^\dagger}}
 \Bigg)\Bigg]~,
\end{equation}
and 
\begin{equation}
\gamma^\mu_\perp=\gamma^\mu-\frac{\nslash}{2}\nbar^\mu
 -\frac{\nbarslash}{2}n^\mu~.
\end{equation}
Here the gluon field is separated into the collinear and soft parts as
$A^\mu=A_c^\mu+A_s^\mu$, and the collinear gluon field $A_n$ is given by
\begin{equation}
A_c^\mu(x)=\sum_{\tilde q}e^{-i{\tilde q}\cdot x}A_{n,q}^\mu(x)~.
\end{equation}
The operator ${\bar\calP}$ acts on the collinear fields as
\begin{eqnarray}
\lefteqn{
f({\bar\calP})(\phi^\dagger_{q_1}\cdots\phi^\dagger_{q_m}
 \phi_{p_1}\cdots\phi_{p_n})}\nonumber\\
&=&
f(\nbar\cdot p_1+\cdots+\nbar\cdot p_n-\nbar\cdot q_1-\cdots-\nbar\cdot q_m)
(\phi^\dagger_{q_1}\cdots\phi^\dagger_{q_m}\phi_{p_1}\cdots\phi_{p_n})~.
\end{eqnarray}
And the heavy quark fields $h_v$ are accompanied in a HQET form,
\begin{equation}
\calL_{\rm HQET}={\bar h}_v iv\cdot Dh_v~.
\end{equation}
\par
The vector current $V^\mu={\bar q}\gamma^\mu b$ is then matched to the effective
currents of the SCET up to $\calO(\lambda)$ as \cite{Chay}
\begin{equation}
V^\mu\to\sum_iC_i(\mu)J_i^\mu+\sum_j B_j(\mu)O_j^\mu+\sum_kA_k(\mu)T_k^\mu~,
\end{equation}
where
\begin{equation}
J_1^\mu={\bar\xi}_nW\gamma^\mu h_v~,~~~
J_2^\mu={\bar\xi}_nWv^\mu h_v~,~~~
J_3^\mu={\bar\xi}_nWn^\mu h_v~,
\end{equation}
\begin{eqnarray}
O_1^\mu&=&{\bar\xi}_n\frac{\nbarslash}{2}(\calPslash_\perp-g\Aslash_\perp)
 W\frac{1}{{\bar\calP^\dagger}}\gamma^\mu h_v~,\nonumber\\
O_2^\mu&=&{\bar\xi}_n\frac{\nbarslash}{2}(\calPslash_\perp-g\Aslash_\perp)
 W\frac{1}{{\bar\calP^\dagger}}v^\mu h_v~,\nonumber\\
O_3^\mu&=&{\bar\xi}_n\frac{\nbarslash}{2}(\calPslash_\perp-g\Aslash_\perp)
 W\frac{1}{{\bar\calP^\dagger}}n^\mu h_v~,\nonumber\\
O_4^\mu&=&{\bar\xi}_n(\calPslash^\mu_\perp-g\Aslash^\mu_\perp)
 W\frac{1}{{\bar\calP^\dagger}}h_v~.
\end{eqnarray}
and
\begin{equation}
T_k^\mu=i\int d^4y T\{J^\mu_k(0),\calL_1(y)\}~~~(k=1,2,3)~.
\end{equation}
The Wilson coefficients $C_i(\mu)$, $B_i(\mu)$, and $A_k(\mu)$ are summarized
in the Appendix A.
\par
Now the weak transition matrix elements of $B\to\pi(\rho)$ are simplified,
introducing nonperturbative functions $\xi_{P,\|,\perp}$ at leading order
and $a(b)_{P,V1,V2}$ at subleading order of $\lambda$ \cite{Chay}:
\begin{mathletters}
\label{SCET}
\begin{eqnarray}
\label{BVpi}
\langle\pi(p)|V^\mu|B(p_B)\rangle&=&
 2En^\mu\left\{(C_1+C_3)\xi_P+\frac{1}{2E}[(-B_1+B_3)a_P+(A_1+A_3)b_P]\right\}
\nonumber\\
&& +2Ev^\mu\left\{C_2\xi_P+\frac{1}{2E}[(2B_1+B_2)a_P+A_2b_P]\right\}~,\\
\label{BVrho}
\langle\rho(p,\epsilon^*)|V^\mu|B(p_B)\rangle&=&
 2Ei\epsilon^{\mu\nu\alpha\beta}\epsilon^*_\nu v_\alpha n_\beta\left\{
 C_1\xi_\perp+\frac{1}{2E}(B_4a_{V1}+A_1b_{V1})\right\}~,\\
\label{BArho}
\langle\rho(p,\epsilon^*)|A^\mu|B(p_B)\rangle&=&
 2E\epsilon^{*\mu}\left\{C_1\xi_\perp+\frac{1}{2E}(B_4a_{V1}+A_1b_{V1})\right\}
\nonumber\\
&& +2E(\epsilon^*\cdot v)v^\mu\left\{
   C_2\xi_\|+\frac{1}{2E}[(2B_1+B_2)a_{V2}+A_2b_{V2}]\right\}\nonumber\\
&&-2E(\epsilon^*\cdot v)n^\mu\Bigg\{C_1\xi_\perp-(C_1+C_3)\xi_\|
 +\frac{1}{2E}[(B_1-B_3)a_{V2} \nonumber\\
&& +B_4a_{V1}+A_1b_{V1}-(A_1+A_3)b_{V2}]\Bigg\}~.
\end{eqnarray}
\end{mathletters}

\section{Light cone sum rules}

Light cone sum rule begins with the two-point (2P) correlation function
\begin{equation}
F^\mu_{B\to\pi}=i\int d^4x e^{iq\cdot x}\langle\pi(p)|T\ubar(x)\gamma^\mu b(x)
 j^+_B(0)|0\rangle~,
\label{B2pi2P}
\end{equation}
where $j^+_B(x)={\bar b}(x)i\gamma_5 d(0)$.
In the phenomenological representation, it simply becomes
\begin{equation}
F^\mu_{B\to\pi}=
 \frac{\langle\pi(p)|\ubar\gamma^\mu b|B\rangle\langle B|j^+_B|0\rangle}
      {m_B^2-(p+q)^2}
 +\sum_{H\ne B}
 \frac{\langle\pi(p)|\ubar\gamma^\mu b|H\rangle\langle H|j^+_B|0\rangle}
      {m_H^2-(p+q)^2}~.
\label{B2pipheno}
\end{equation}
The second term of (\ref{B2pipheno}) corresponds to the resonance part.
In the HQET and SCET, the matrix elements of the first term of 
(\ref{B2pipheno}) are parametrized as in (\ref{BVpi}) and
\begin{equation}
\langle 0|\qbar\Gamma h^{(b)}_v|B_v\rangle=\frac{F}{2}Tr[\Gamma\calM_v]~,
\end{equation}
where
\begin{equation}
\calM_v=\frac{1+\vslash}{2}(-\gamma_5)~.\nonumber\\
\end{equation}
Here $F$ is the $B$ meson decay constant in the effective theory.
\par
To establish the light cone sum rule, one needs to evaluate (\ref{B2pi2P})
in terms of the pion distribution amplitudes (up to twist 4),
\begin{mathletters}
\label{piDA}
\begin{eqnarray}
\langle\pi(p)|\ubar(x)\gamma^\mu\gamma_5 d(0)|0\rangle&=&
 -ip^\mu f_\pi\int_0^1 du~ e^{iup\cdot x}[\phi_\pi(u)+x^2g_1(u)]\nonumber\\
 &&+f_\pi\left(x^\mu-\frac{x^2p^\mu}{x\cdot p}\right)
   \int_0^1 du~ e^{iup\cdot x}g_2(u)~,\\
\langle\pi(p)|\ubar(x)i\gamma_5 d(0)|0\rangle&=&
 f_\pi\mu_\pi\int_0^1 du~ e^{iup\cdot x}\phi_p(u)~,\\
\langle\pi(p)|\ubar(x)\sigma^{\mu\nu}\gamma_5 d(0)|0\rangle&=&
 if_\pi\frac{\mu_\pi}{6}(p^\mu x^\nu-p^\nu x^\mu)
  \int_0^1 du~ e^{iup\cdot x}\phi_\sigma(u)~,
\end{eqnarray}
\end{mathletters}
where $\mu_\pi\equiv m_\pi^2/(m_u+m_d)$.
\par
After a proper Borel transformation ($T$ is the associated Borel parameter),
\begin{eqnarray}
&&\frac{iFe^{-2\Lambdabar/T}}{m_b}\left[
2En^\mu\left\{(C_1+C_3)\xi_P+\frac{1}{2E}[(-B_1+B_3)a_P+(A_1+A_3)b_P]\right\}
\right.\nonumber\\
&&\left.
+2Ev^\mu\left\{C_2\xi_P+\frac{1}{2E}[(2B_1+B_2)a_P+A_2b_P]\right\}\right]
\nonumber\\
&=&\frac{1}{\pi}\int_0^{s_0}{\rm Im}F^\mu_{B\to\pi}(E,s)e^{-s/T}ds~.
\label{sumruleB2pi}
\end{eqnarray}
The imaginary part of $F^\mu_{B\to\pi}$ is 
\begin{eqnarray}
\frac{1}{\pi}F^\mu_{B\to\pi}(E,s)&=&
f_\pi\Theta(\ubar_0)v^\mu\Bigg[\frac{\mu_\pi}{2E}\phi_p(\ubar_0)
 -\frac{g'_2(\ubar_0)}{2E^2}+\frac{1}{2E}\frac{\mu_\pi}{6}\phi'_\sigma(\ubar_0)
\Bigg]\nonumber\\
&&
+f_\pi\Theta(\ubar_0)p^\mu\Bigg[\frac{\phi_\pi(\ubar_0)}{2E}
 -\frac{1}{2E^2}\frac{\mu_\pi}{6}\phi'_\sigma(\ubar_0)
 +\frac{g'_2(\ubar_0)-g''_1(\ubar_0)}{2E^3}\Bigg]~,
\end{eqnarray}
where $\ubar_0=1-s/2E$.
\par
Comparing both sides of (\ref{sumruleB2pi}) gives the final results
\begin{mathletters}
\label{AB}
\begin{eqnarray}
\label{A}
\lefteqn{
2E\left\{C_2\xi_P+\frac{1}{2E}[(2B_1+B_2)a_P+A_2b_P]\right\}}\nonumber\\
&=&-if_\pi N\left\{
 \int_0^\theta du\left[\mu_\pi\phi_p(\ubar)+\frac{2}{T}g_2(\ubar)
 -\frac{2E}{T}\frac{\mu_\pi}{6}\phi_\sigma(\ubar)\right]e^{-2uE/T}
\right.\nonumber\\
&&\left.
 +\left[\frac{1}{E}g_2({\bar\theta})
   -\frac{\mu_\pi}{6}\phi_\sigma({\bar\theta})\right]e^{-2\theta E/T}\right\}
~,\\
\label{B}
\lefteqn{
2E\left\{(C_1+C_3)\xi_P+\frac{1}{2E}[(-B_1+B_3)a_P+(A_1+A_3)b_P]\right\}}
\nonumber\\
&=&-if_\pi N\left\{
 \int_0^\theta du\left[\phi_\pi(\ubar)
   +\frac{2}{T}\frac{\mu_\pi}{6}\phi_\sigma(\ubar)-\frac{2}{2ET}g_2(\ubar)
   -\frac{4}{T^2}g_1(\ubar)\right]e^{-2uE/T}\right.\nonumber\\
 &&\left.+\left[\frac{1}{E}\frac{\mu_\pi}{6}\phi_\sigma({\bar\theta})
  -\frac{1}{E^2}g_2({\bar\theta})+\frac{1}{E^2}g'_1({\bar\theta})
  -\frac{2}{ET}g_1({\bar\theta})\right]e^{-2\theta E/T}\right\}~,
\end{eqnarray}
\end{mathletters}
where $\ubar=1-u$, $\theta={\rm min}(1, s_0/2E)$, $\bar\theta=1-\theta$, and 
$N=m_b e^{2\Lambdabar/T}/F$.
\par
For the vector meson production, we calculate the 2P function
\begin{eqnarray}
\label{B2rho2P}
F_{B\to\rho}^\mu&=&i\int d^4x~e^{-ip_B\cdot x}\langle\rho(p,\epsilon^*)|
 T \ubar(0)\gamma^\mu(1-\gamma_5)b(0)j^+_B(x)|0\rangle\nonumber\\
&=&
 \frac{\langle\rho(p,\epsilon^*)|\ubar\gamma^\mu(1-\gamma_5) b|B\rangle\langle B|j^+_B|0\rangle}
      {m_B^2-(p+q)^2}
 +\sum_{H\ne B}
 \frac{\langle\rho(p,\epsilon^*)|\ubar\gamma^\mu(1-\gamma_5) b|H\rangle\langle H|j^+_B|0\rangle}
      {m_H^2-(p+q)^2}~,
\end{eqnarray}
where the second line is the phenomenological description.
\par
Using (\ref{SCET}), the sum rule has the form of 
\begin{eqnarray}
&&
\frac{iFe^{-2\Lambdabar/T}}{m_b}\left[
2E\epsilon^{*\mu}\left\{C_1\xi_\perp+\frac{1}{2E}(B_4a_{V1}+A_1b_{V1})\right\}
\right.\nonumber\\
&& +2E(\epsilon^*\cdot v)v^\mu\left\{
   C_2\xi_\|+\frac{1}{2E}[(2B_1+B_2)a_{V2}+A_2b_{V2}]\right\}\nonumber\\
&&-2E(\epsilon^*\cdot v)n^\mu\left\{C_1\xi_\perp-(C_1+C_3)\xi_\|
 +\frac{1}{2E}[(B_1-B_3)a_{V2}\right. \nonumber\\
&&\left.
 +B_4a_{V1}+A_1b_{V1}-(A_1+A_3)b_{V2}]\Bigg\}\right]
=\frac{1}{\pi}\int_0^{s_0}{\rm Im}F^\mu_{B\to\rho}(E,s)e^{-s/T}ds~.
\label{sumruleB2rho}
\end{eqnarray}
Introducing the distribution amplitudes of $\rho(p,\epsilon^*)$ (up to twist 3)
\begin{mathletters}
\label{rhoDA}
\begin{eqnarray}
\langle\rho(p,\epsilon^*)|\ubar(0)\sigma^{\mu\nu}d(x)|0\rangle&=&
-if_\rho^\perp(\epsilon^{*\mu}p^\nu-\epsilon^{*\nu}p^\mu)
 \int_0^1du~e^{iup\cdot x}\phi_\perp(u)~,\\
\langle\rho(p,\epsilon^*)|\ubar(0)\gamma^\mu d(x)|0\rangle&=&
f_\rho m_\rho p^\mu\frac{\epsilon^*\cdot x}{p\cdot x}
 \int_0^1 du~e^{iup\cdot x}\phi_\|(u)\nonumber\\
&&
+f_\rho m_\rho\left(
 \epsilon^{*\mu}- p^\mu\frac{\epsilon^*\cdot x}{p\cdot x}\right)
 \int_0^1 du~e^{iup\cdot x}g_\perp^{(v)}(u)~,\\
\langle\rho(p,\epsilon^*)|\ubar(0)\gamma^\mu\gamma_5 d(x)|0\rangle&=&
\frac{1}{4}f_\rho m_\rho\epsilon^{\mu\nu\alpha\beta}\epsilon^*_\nu p_\alpha
x_\beta \int_0^1 du~e^{iup\cdot x}g_\perp^{(a)}(u)~,
\end{eqnarray}
\end{mathletters}
the imaginary part of 2P function becomes
\begin{eqnarray}
\frac{1}{\pi}{\rm Im}F^\mu_{B\to\rho}(E,s)&=&\frac{\Theta(\ubar_0)}{2iE}\left[
-\frac{1}{4}f_\rho m_\rho\epsilon^{\mu\nu\alpha\beta}\epsilon^*_\nu p_\alpha
 v_\beta\frac{i}{E}g^{(a)\prime}_\perp(u_0)
+if_\rho^\perp\epsilon^{\mu\nu\alpha\beta}v_\nu\epsilon^*_\alpha p_\beta
 \phi_\perp(u_0)\right.\nonumber\\
&&
f_\rho m_\rho p^\mu\frac{\epsilon^*\cdot v}{E}\phi_\|(u_0)
+f_\rho m_\rho\left(\epsilon^{*\mu}-p^\mu\frac{\epsilon^*\cdot v}{E}\right)
 g^{(v)}_\perp(u_0)\nonumber\\
&&\left.
+v_\nu f_\rho^\perp(\epsilon^{*\mu}p^\nu-\epsilon^{*\nu}p^\mu)\phi_\perp(u_0)
\right]~,
\label{ImFrho}
\end{eqnarray}
where $u_0\equiv s/2E$.
Plugging (\ref{ImFrho}) into (\ref{sumruleB2rho}), we arrive at
\begin{mathletters}
\label{finalB2rho}
\begin{eqnarray}
\label{e}
&2E&\left\{C_1\xi_\perp+\frac{1}{2E}(B_4a_{V1}+A_1b_{V1})\right\}=
-\frac{N}{2E}\int_0^\kappa ds~e^{-s/T}\Bigg[
 \frac{1}{4}f_\rho m_\rho g^{(a)\prime}_\perp(u_0)-Ef_\rho^\perp\phi_\perp(u_0)
 \Bigg]~,\nonumber\\
&&\\
\label{estar}
&2E&\left\{C_1\xi_\perp+\frac{1}{2E}(B_4a_{V1}+A_1b_{V1})\right\}=
\frac{N}{2E}\int_0^\kappa ds~e^{-s/T}\Big[f_\rho m_\rho g^{(v)}_\perp(u_0)
 +Ef_\rho^\perp\phi_\perp(u_0)\Big]~,\\
\label{v}
&2E&\left\{C_2\xi_\|+\frac{1}{2E}[(2B_1+B_2)a_{V2}+A_2b_{V2}]\right\}=0~,\\
\label{n}
&2E&\left\{C_1\xi_\perp-(C_1+C_3)\xi_\|+\frac{1}{2E}[(B_1-B_3)a_{V2}
 +B_4a_{V1}+A_1b_{V1}-(A_1+A_3)b_{V2}]\right\}\nonumber\\
&=&
-\frac{N}{2E}\int_0^\kappa ds~e^{-s/T}\Big[f_\rho m_\rho\{\phi_\|(u_0)
 -g^{(v)}_\perp(u_0)\}-Ef_\rho^\perp\phi_\perp(u_0)\Big]~.
\end{eqnarray}
\end{mathletters}

\section{Energy scalings of the form factors}

At leading order of $\lambda$, SCET introduces three independent functions
$\xi_P$, $\xi_\perp$, and $\xi_\|$ to describe $B\to\pi(\rho)$ transition.
At the NLO, additional six subleading functions $a_i$ and $b_i$ ($i=P,V1,V2$)
are needed.
We can now evaluate these functions from the LCSR established in (\ref{AB}) and
(\ref{finalB2rho}).
\par
Let us see first Eq.\ (\ref{AB}).
For sufficiently large $E$ and $m_b$ (at leading order of $\alpha_s$)
\begin{mathletters}
\begin{eqnarray}
\label{aP}
&2a_P&=-if_\pi N\Bigg\{
 \mu_\pi({\tilde B}_2+{\tilde B}_4-6{\tilde C}_2-15{\tilde C}_4)\frac{T}{2E}
 +\Bigg[\frac{20\delta^2}{3T}-\mu_\pi(6{\tilde B}_2+20{\tilde B}_4)\Bigg]
   \frac{T^2}{4E^2}\Bigg\}~,\\
\label{xiP}
\lefteqn{
2E\Bigg\{\xi_P+\frac{-a_P+b_P}{2E}\Bigg\}}\nonumber\\
&=&-if_\pi N\frac{1}{4E}\Bigg\{\Big[
 6(1+6a_2+15a_4)T^2+2\mu_\pi(1+6{\tilde C}_2+15{\tilde C}_4)T-4\epsilon\delta^2
 \Big]-\frac{40\delta^2}{3}\frac{T}{2E}\Bigg\}~.
\end{eqnarray}
\end{mathletters}
From (\ref{xiP}), the energy scaling of leading and subleading functions are
\begin{equation}
\xi_P+\frac{-a_P+b_P}{E}\sim\frac{1}{E^2}+\frac{1}{E^3}~.
\label{scaling}
\end{equation}
Such a behavior is quite reasonable and very compatible with the previously 
known $\xi_P\sim 1/E^2$.
Referring to (\ref{scaling}), Eq.\ (\ref{aP}) implies
\begin{equation}
\label{BC}
{\tilde B}_2+{\tilde B_4}-6{\tilde C}_2-15{\tilde C}_4 =0~,
\end{equation}
which corresponds to (see (\ref{phip}) and (\ref{phisigma}))
\begin{equation}
\phi_p(1)+\frac{1}{6}\phi'_\sigma(1)=0~,
\end{equation}
of \cite{Charles}.
Numerically,
\begin{equation}
{\tilde B}_2+{\tilde B}_4-6{\tilde C}_2-15{\tilde C}_4=0.025~,
\end{equation}
where the values of (\ref{number}) are used.
\par
From the sum rules for $B\to\rho$ decay, we have, by (\ref{e}) and(\ref{estar}),
\begin{equation}
\int_0^\theta ds e^{-s/T}\Bigg[
 g^{(v)}_\perp(u_0)+\frac{1}{4}g^{(a)\prime}_\perp(u_0)\Bigg]=0~.
\end{equation}
It should be understood that the relation holds near the end-point region.
At $u=1$, the relation is exact from (\ref{gv}) and (\ref{ga}),
\begin{equation}
g^{(v)}_\perp(u)+\frac{1}{4}g^{(a)\prime}_\perp(u)=
\int_u^1 dv \frac{\phi_\|(v)}{v}~.
\end{equation}
In the vicinity of $u\sim 1$, we use
\begin{eqnarray}
g^{(v)}_\perp(u)&\approx&\frac{3}{2}(1+a_2^\|)~,\nonumber\\
g^{(a)}_\perp(u)&\approx&6(1+a_2^\|)\ubar~,
\end{eqnarray}
to calculate (\ref{e})
\begin{eqnarray}
\lefteqn{
\frac{(2E)^2}{N}\Bigg\{\xi_\perp+\frac{b_{V1}}{2E}\Bigg\}}\nonumber\\
&=&\Bigg[f_\rho m_\rho\frac{3}{2}(1+a_2^\|)T+f_\rho^\perp(3+18a_2^\|)T^2\Bigg]
 -\Bigg[f_\rho m_\rho(3+18a_2^\|)T^2+f_\rho^\perp(6+216a_2^\perp)T^3\Bigg]
   \frac{1}{2E}~.\nonumber\\
\end{eqnarray}
Thus we have
\begin{equation}
\xi_\perp+\frac{b_{V1}}{E}\sim\frac{1}{E^2}+\frac{1}{E^3}~.
\label{perp}
\end{equation}
Eq.\ (\ref{v}) tells that $a_{V2}=0$.
The absence of $v^\mu$ component in the sum rules is due to the twist expansion
in terms of the $\rho$ meson's distribution amplitudes (\ref{rhoDA}).
Vanishing $a_{V2}$ does not mean that it is suppressed by $\lambda^n$.
Instead, at higher twist, there are terms proportional to $x^\mu$ (which turns 
out to be proportional to $v^\mu$ because the heavy quark propagator is 
$\langle T h_v(0) {\bar h}_v(x)\rangle
=\int_0^\infty dt\delta^4(-x-vt)(1+\vslash)/2$)
in (\ref{rhoDA}) with new distribution functions \cite{Ball}.
\par
Another nontrivial relation comes out from (\ref{estar})$-$(\ref{n}):
\begin{eqnarray}
\frac{(2E)^2}{N}\Bigg\{\xi_\|+\frac{-a_{V2}+b_{V2}}{2E}\Bigg\}
&=&f_\rho m_\rho\int_0^\theta ds ~e^{-s/T}\phi_\|(u_0)\nonumber\\
&=&f_\rho m_\rho\Bigg[(6+36a_2^\|)\frac{T^2}{2E}-(6+216a_2^\|)\frac{2T^3}{4E^2}
 +\cdots\Bigg]~.
\end{eqnarray}
Considering (\ref{v}), the above equation means that
\begin{equation}
\xi_\|+\frac{b_{V2}}{E}\sim\frac{1}{E^3}+\frac{1}{E^4}~.
\label{parallel}
\end{equation}
The longitudinal component is, therefore, $1/E$-suppressed compared to the 
transverse one of (\ref{perp}).
\par
Some remarks are in order.
First, the LCSR calculation is done at leading order of $\alpha_s$.
In \cite{Kho,Ball2}, leading radiative corrections to the twist-2,3 
contributions of $B\to\pi$ are calculated.
Corrections to the twist-2 are rather large ($\approx 30\%$), but they are
almost canceled by large radiative corrections to the decay constant $f_B$,
leaving the form factor $f_+$ free from large corrections \cite{Kho}.
Authors of \cite{Ball2} found that the sizes of the radiative corrections to
the twist-3 distribution amplitudes $\phi_p$ and $\phi_\sigma$ are of the
same size as that of the twist-2 while the sum of both is quite small.
Hence it would be very interesting to check whether that kind of 
{\em cancellation} might occur at higher twist and in $B\to\rho$ decays.
One more point to be noticed is that the radiative corrections preserve the
factorized scheme of the LCSR \cite{Ball2},
\begin{equation}
({\rm 2P~functions})\sim\sum_{\rm twists} T_H\otimes\phi~,
\label{factorization}
\end{equation}
where $T_H$ is the process-dependent hard kernel and $\phi$ is the 
nonperturbative distribution amplitudes, up to twist-3.
Such a factorization in (\ref{factorization}) is desirable to establish
sum rules between the LCSR and the SCET.
The role of $T_H$ is very similar to that of the Wilson coefficients in SCET
which separate the long- and short-distance physics.
Thus the study of short-distance structure will check the validity of the sum
rule, or provide more information on both sides.
\par
Second, we have neglected the hard contributions.
Hard contributions arise when an energetic light quark from the weak vertex 
exchanges the hard gluon with the soft degrees of freedom.
But it was argued that the effects are suppressed by 
$\alpha_s(\sqrt{m_b\Lambda_{\rm QCD}})$ \cite{Beneke}.
Also, explicit calculations in \cite{Kho,Ball2} show that the soft contributions
which are responsible for the case where the recoiling quark carries almost
all the momentum are dominant.
\footnote{Recent work of \cite{Hill} includes the hard contribution to extend 
the SCET.}
\par

\section{Summary}

In this paper we established the light-cone sum rules for the $B\to\pi(\rho)$
form factors in the context of SCET.
Both LCSR and SCET are useful tools to describe the heavy-to-light transitions
at large recoil, and favor the asymmetric configuration of quark and antiquark
to form the energetic meson through the "soft contributions".
From the sum rules,
leading and subleading nonperturbative functions at $\calO(\lambda)$ of SCET
are found to be $\sim 1/E^2+1/E^3$, which is compatible with other literatures.
The fact that radiative corrections to the LCSR preserve the separation between
long- and short-distance physics is encourageous to study the hybrid scheme of
SCET and LCSR.
More refinements in every direction such as radiative corrections, hard
spectator effects, etc. will provide better understanding of $B\to\pi(\rho)$ as
well as $|V_{ub}|$.

\begin{center}
{\large\bf Acknowledgements}\\[10mm]
\end{center}

This work was supported by the BK21 Program of the Korean Ministry of Education.

\begin{appendix}

\section{Wilson coefficients}

In this Appendix, Wilson coefficients for the SCET weak currents are given 
\cite{Chay}:
\begin{eqnarray}
C_1(\mu)&=&1-\frac{\alpha_s C_F}{4\pi}\Bigg[2\ln^2\Bigg(\frac{xm_b}{\mu}\Bigg)
 -5\ln\frac{m_b}{\mu}+\frac{3x-2}{1-x}\ln x+2Li_2(1-x)+\frac{\pi^2}{12}+6\Bigg]
 \nonumber\\
C_2(\mu)&=&\frac{\alpha_s C_F}{4\pi}\Bigg[\frac{2}{1-x}+\frac{2x}{(1-x)^2}\ln x
 \Bigg]\nonumber\\
C_3(\mu)&=&\frac{\alpha_s C_F}{4\pi}\Bigg[-\frac{x}{1-x}
 +\frac{x(1-2x)}{(1-x)^2}\ln x\Bigg]~,
\end{eqnarray}
where $x=2E/m_b$, $C_F$ is the color factor, and 
$Li_2(x)=-\int_0^x\frac{dt}{t}\ln(1-t)$ is the dilogarithmic function.
Other coefficients are related to $C_i$ as
\begin{eqnarray}
B_i(\mu)&=&C_i(\mu)~(i=1,2,3)~,~~~B_4(\mu)=2C_3(\mu)~,\nonumber\\
A_i(\mu)&=&C_i(\mu)~.
\end{eqnarray}

\section{Distribution amplitudes}

We summarize the distribution amplitudes for $\pi(p)$ and $\rho(p,\epsilon^*)$:
\begin{mathletters}
\begin{eqnarray}
\label{phipi}
\phi_\pi(u,\mu)&=&6u\ubar\Bigg\{1+a_2(\mu)\frac{3}{2}[5(u-\ubar)^2-1]
 +a_4(\mu)\frac{15}{8}[21(u-\ubar)^4-14(u-\ubar)^2+1]\Bigg\}~,\\
\label{phip}
\phi_p(u,\mu)&=&1+{\tilde B}_2(\mu)\frac{1}{2}[3(u-\ubar)^2-1]
 +{\tilde B}_4(\mu)\frac{1}{8}[35(u-\ubar)4-30(u-\ubar)^2+3]~,\\
\label{phisigma}
\phi_\sigma(u,\mu)&=&6u\ubar\Bigg\{
 1+{\tilde C}_2(\mu)\frac{3}{2}[5(u-\ubar)^2-1]
 +{\tilde C}_4(\mu)\frac{15}{8}[21(u-\ubar)^4-14(u-\ubar)^2+1]\Bigg\}~,\\
\label{g1}
g_1(u,\mu)&=&\frac{5}{2}\delta^2(\mu)u^2\ubar^2
 +\frac{1}{2}\epsilon(\mu)\delta^2(\mu)\Bigg\{
 u\ubar(2+13u\ubar)+10u^3\ln u\Bigg(2-3u+\frac{6}{5}u^2\Bigg)\nonumber\\
 &&+10\ubar^3\ln \ubar\Bigg(2-3\ubar+\frac{6}{5}\ubar^2\Bigg)~,\\
\label{g2}
g_2(u,\mu)&=&\frac{10}{3}\delta^2(\mu)u\ubar(u-\ubar)~,
\end{eqnarray}
\end{mathletters}
for the pion, and
\begin{mathletters}
\begin{eqnarray}
\phi_{\perp(\|)}(u,\perp)&=&6u\ubar\Bigg\{1+a_2^{\perp(\|)}\frac{3}{2}
 [5(u-\ubar)^2-1]\Bigg\}~,\\
\label{gv}
g^{(v)}_\perp(u,\mu)&=&\frac{1}{2}\Bigg[
 \int_0^u dv\frac{\phi_\|(v,\mu)}{1-v}+\int_u^1\frac{\phi_\|(v,\mu)}{v}\Bigg]
~,\\
\label{ga}
g^{(a)}_\perp(u,\mu)&=&2\Bigg[(1-u)\int_0^u dv\frac{\phi_\|(v,\mu)}{1-v}
 +u\int_u^1\frac{\phi_\|(v,\mu)}{v}\Bigg]~,
\end{eqnarray}
\end{mathletters}
for $\rho$ meson.
Here $\mu$ is the renormalization scale. 
The values of the coefficients at $\mu=1$ GeV are
\begin{eqnarray}
\label{number}
a_2&=&0.44~,~~~a_4=0.25~,~~~{\tilde B}_2=0.48~,~~~{\tilde B}_4=1.15~,\nonumber\\
{\tilde C}_2&=&0.10~,~~~{\tilde C}_4=0.067~,~~~\delta^2=0.2~{\rm GeV}^2~,~~~
\epsilon=0.5~,\nonumber\\
a_2^\perp&=&0.2~,~~~a_2^\|=0.18~.
\end{eqnarray}

\end{appendix}


\begin{thebibliography}{99}

\bibitem{BABAR}
B.\ Aubert \etal, the $BABAR$ Collaboration, hep-ex/0301001.
\bibitem{HQET}
N.\ Isgur and M.B.\ Wise, \plb{232}{1989}{113}; 
\ibid{237}{1990}{527}.
\bibitem{Vcb}
M.\ Luke, \plb{252}{1990}{447};
A.F.\ Falk, H.\ Georgi, B.\ Grinstein, and M.B.\ Wise, \npb{343}{1990}{1};
M.\ Neubert, \plb{264}{1991}{455}.
\bibitem{LEET}
M.J.\ Dugan and B.\ Grinstein, \plb{255}{1991}{583}.
\bibitem{SCET}
C.W.\ Bauer, S.\ Fleming, and M.\ Luke, \prd{63}{2000}{014006};
C.W.\ Bauer, S.\ Fleming, D.\ Pirjol, and I.W.\ Stewart, \prd{63}{2001}{114020}.
\bibitem{Chay}
J.\ Chay and C.\ Kim, \prd{65}{2002}{114016}.
\bibitem{Charles}
J.\ Charles, A.\ Le Yaouanc, L.\ Oliver, O.\ P\`ene, and J.-C.\ Raynal, 
\prd{60}{1999}{014001}.
\bibitem{Ball0}
P.\ Ball and V.M.\ Braun, \prd{55}{1997}{5561}.
\bibitem{Chernyak}
V.L.\ Chernyak and I.R.\ Zhitnitsky, \npb{345}{1990}{137}.
\bibitem{Ali}
A.\ Ali, V.M.\ Braun, and H.\ Simma, \zpc{63}{1994}{437};
P.\ Ball and V.M.\ Braun, \prd{54}{1996}{2182}.
\bibitem{Wang}
W.Y.\ Wang and Y.L.\ Wu, \plb{515}{2001}{57}; 
\ibid{519}{2001}{219}.
\bibitem{Liu}
J.G.\ K\"orner, C.\ Liu, and C.-T.\ Yan, \prd{66}{2002}{076007}.
\bibitem{Ball}
P.\ Ball, V.M.\ Braun, Y.\ Koike, and K.\ Tanaka, \npb{529}{1998}{323}.
\bibitem{Kho}
A.\ Khodjamirian, R.\ R\"uckl, S.\ Weinzierl, and O.\ Yakovlev,
\plb{410}{1997}{275};
E.\ Bagan, P.\ Ball, and V.M.\ Braun, \plb{417}{1998}{154}.
\bibitem{Ball2}
P.\ Ball and R.\ Zwicky, \jhep{0110}{2001}{019}.
\bibitem{Beneke}
M.\ Beneke and Th. Feldmann, \npb{592}{2001}{3}.
\bibitem{Hill}
R.J.\ Hill and M.\ Neubert, hep-ph/0211018.

\end{thebibliography}
\end{document}